# Wide operational windows of edge-localized mode suppression by resonant magnetic perturbations in the DIII-D tokamak

Q.M. Hu[1], R. Nazikian[1], B.A. Grierson[1], N.C. Logan[1], D. M. Orlov[2], C. Paz-Soldan[3], and Q. Yu[4]

[1]*Princeton Plasma Physics Laboratory, Princeton NJ 08543-0451, USA*
[2]*University of California, San Diego, 9500 Gilman Drive, La Jolla, California 92093-0417, USA*
[3]*General Atomics, PO Box 85608, San Diego, California 92186-5608, USA*
[4]*Max-Plank-Institut fur Plasmaphysik, 85748 Garching, Germany*

Edge-Localized-Mode (ELM) suppression by resonant magnetic perturbations (RMPs) generally occurs over very narrow ranges of the plasma current (or magnetic safety factor $q_{95}$) in the DIII-D tokamak. However, wide $q_{95}$ ranges of ELM suppression are needed for the safety and operational flexibility of ITER and future reactors. In DIII-D ITER Similar Shape (ISS) plasmas with $n=3$ RMPs, the range of $q_{95}$ for ELM suppression is found to increase with decreasing electron density. Nonlinear two-fluid MHD simulations reproduce the observed $q_{95}$ windows of ELM suppression and the dependence on plasma density, based on the conditions for resonant field penetration at the top of the pedestal. When the RMP amplitude is close to the threshold for resonant field penetration, only narrow isolated magnetic islands form near the top of the pedestal, leading to narrow $q_{95}$ windows of ELM suppression. However, as the threshold for field penetration decreases with decreasing density, resonant field penetration can take place over a wider range of $q_{95}$. For sufficiently low density (penetration threshold) multiple magnetic islands form near the top of the pedestal giving rise to continuous $q_{95}$ windows of ELM suppression. The model predicts that wide $q_{95}$ windows of ELM suppression can be achieved at substantially higher pedestal pressure in DIII-D by shifting to higher toroidal mode number ($n=4$) RMPs.

Magnetically confined high temperature fusion plasmas exhibit spontaneous transitions to high thermal confinement called H-mode [1], a requirement for fusion reactors to produce net electricity. The H-mode is characterized by the suppression of edge turbulence concomitant with the formation of a steep edge pressure gradient, known as the pedestal [1,2]. Unfortunately, the H-mode is also subject to magnetohydrodynamic instabilities called Edge-Localized-Modes (ELMs) that can release damaging bursts of heat and particles to the reactor chamber. A goal of fusion research is to find ways of suppressing ELMs while maintaining favourable H-mode confinement. ELM suppression has been achieved using weak 3D helical magnetic fields (known as resonant magnetic perturbations - RMPs) in present experiments [3], and there are plans to control ELMs by similar RMPs in ITER [4]. Currently ELM suppression has been achieved using a variety of RMPs in a range of fusion experiments worldwide [3,5–17], however a key question is whether ITER can access high fusion power operation consistent with the requirements for ELM suppression.

Operating over a wide range of operating parameter such as the plasma current (or alternatively the magnetic safety factor $q_{95}$) is an important requirement for existing and future fusion experiments, however RMPs have only achieved narrow $q_{95}$ windows of ELM suppression with rare exception [6–8,13]. Some progress has been made in understanding the role of magnetic islands on the $q_{95}$ windows of ELM suppression based vacuum field analysis [7,18]. However, such calculations typically produce extensive regions of magnetic stochasticity that are inconsistent with the high edge temperatures characteristic of H-mode [19,20]. It is now understood that models must take into account the screening response [11,21–23] of high temperature fusion plasmas and the associated transport [24–26] in order to establish consistency with experimental observations. Here we develop a plasma response model based on nonlinear two-fluid MHD simulations [27] that quantitatively accounts for the observed $q_{95}$ windows of ELM suppression and reproduces their dependence on plasma parameters critical for penetration and screening of resonant fields.

Using nonlinear two-fluid MHD simulations, it was recently shown that the observed density and E×B rotational dependence of ELM suppression and density pump-out in DIII-D ITER Similar Shape (ISS) plasmas can be understood from the conditions required for resonant field penetration and the resulting collisional transport. [20,28]. These simulations, based on the TM1 code, used experimental profiles and plasma parameters taken before the RMP was applied, and predicted the profile evolution resulting from the penetration and screening of resonant fields. The results strongly support the early hypothesis that magnetic islands suppress ELMs by limiting the expansion of the pedestal to an unstable width for Peeling-Ballooning-Modes (PBMs) [29]. Here we extend this model to show that wide $q_{95}$ windows of ELM suppression occur when multiple magnetic islands are driven on adjacent rational surfaces $q=m/n$, $m+1/n$ at the top (or knee) of the pedestal.

Here we investigate nominally identical DIII-D ISS plasmas that differ predominantly in the electron density $n_e$. All other parameters remain similar including: toroidal field $B_T$=-1.9T, normalized beta $\beta_N$~1.8-2.2, pedestal temperature $T_{e,ped}$≈1keV, triangularity $\delta$~0.5, neutral beam power 6-8.5MW, and neutral beam torque 5-7.5Nm in the co-$I_p$ direction. The E×B frequency at the pedestal top ($\omega_E=E_r/|RB_\theta|$) is close to -10krad/s for all these plasmas. The plasma current $I_P$ is slowly ramped up over a 2 s interval to scan the magnetic safety factor in each discharge ($q_{95}$≈4.0-3.0). The main shot-to-shot variation is in the pedestal density which varies over the range $n_{e,ped}$≈ 1.5-3.5×10$^{19}$m$^{-3}$, corresponding to $n_{e,ped}/n_G$~0.15-0.35 and pedestal collisionality range $\nu_e^*$≈0.05-0.2. The $n = 3$ RMP is produced using in-vessel coils (I-coils) operating in even parity [30] and the I-coil current is similar (≈4kA) for all these discharges. Figure 1 shows the ELMing behaviour ($D_\alpha$ light in Fig. 1a-c) vs $q_{95}$ for three discharges with different pedestal density. At the highest density (Fig. 1a, $n_{e,ped} \approx 3.5 \times 10^{19}$m$^{-3}$) the $q_{95}$ windows of ELM suppression are very narrow ($\Delta q_{95}$≈0.1) and are aligned with rational surfaces near the top of the pedestal ($q_{95}$≈11/3, 10/3, 9/3). However, as the density decreases (Fig. 1b, $n_{e,ped}$≈2.5×10$^{19}$m$^{-3}$) the ELM suppression windows expand and begin to merge. At the lowest density (Fig. 1c, $n_{e,ped}$≈1.5×10$^{19}$m$^{-3}$) a continuous window of ELM suppression is observed ($\Delta q_{95}$>0.7). The calculated RMP amplitude using GPEC [31] remains similar for all these discharges. A few isolated ELMs can still be observed at low density (Fig. 1c) but these are predominantly triggered by sawteeth. Attaining wide $q_{95}$ windows of ELM suppression is rare in DIII-D as it requires very good wall conditions to achieve the required low density.

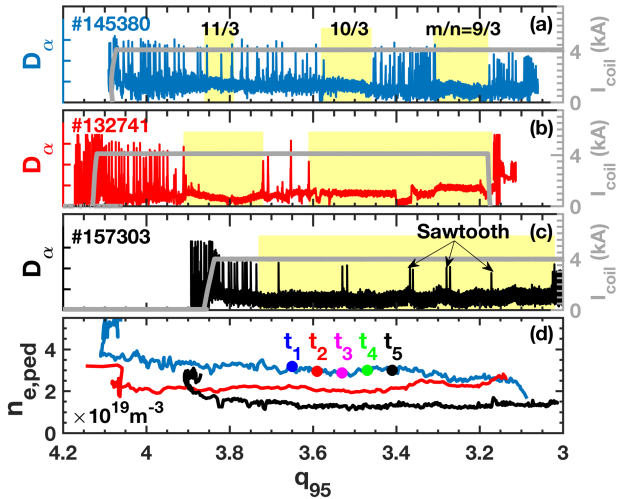

**Fig. 1**. The evolution of (a-c) log of Lyman alpha $D_\alpha$ signals and $n = 3$ I-coil current for three plasmas of varying pedestal density vs $q_{95}$, (d) electron pedestal density $n_{e,ped}$ for #145380 (blue), #132741 (red) and #157303 (black). The windows of ELM suppression are shaded in yellow.

It has been shown from nonlinear two-fluid MHD simulations that the top of pedestal density and E×B frequency *prior* to ELM suppression determine the resonant field penetration threshold ($\delta B_{th}/B_T$), where $\delta B_{th}$ is the minimum RMP amplitude required to drive magnetic islands at the pedestal top (See Fig.16 in Ref. [20]). Based on the density dependence of field penetration (see Eq. 2 in Ref. [20]), we estimate that $\delta B_{th}/B_T$ decreases by ≈ 50% going from high to low pedestal density in Fig. 1. However, the $n=3$ RMP amplitude remains similar for all three discharges. We show that the substantial decrease in the threshold for resonant field penetration with density can account for the expansion of the ELM suppression windows seen in experiment.

At the highest plasma density (Fig. 1a) we observe two narrow windows of ELM suppression, $3.46 < q_{95} < 3.57$ and $3.18 < q_{95} < 3.31$, and a third window of sparse ELMs (not fully suppressed) for $3.8 < q_{95} < 3.85$. These windows are close to low order rational surfaces at the pedestal top, ($q$=11/3 for the sparse ELMing window, and $q$=10/3 and 9/3 for the two ELM suppression windows). At lower density (Fig. 1b) the two ELM suppression windows at $q_{95}$≈10/3 and 9/3 expand and begin to merge into one ELM suppression window ($3.18 < q_{95} < 3.6$). Interestingly, the sparse ELMing window in Fig. 1a for $q_{95}$≈11/3 now becomes fully suppressed ($3.73 < q_{95} < 3.9$) in Fig. 1b. At the lowest pedestal density (Fig. 1c) we obtain one continuous ELM suppression window for $3 < q_{95} < 3.75$. These three discharges are representative of a set of 50 discharges showing the trend of expanding $q_{95}$ windows of ELM suppression with decreasing density.

We use the TM1 code [27,32] to predict the threshold for resonant field penetration and the resulting profile changes due to collisional parallel transport across the magnetic islands. TM1 is a nonlinear two-fluid MHD model in cylindrical geometry [27,32], and includes the diamagnetic drift and ion polarization current that are important for calculating the threshold for resonant field penetration in the pedestal. The TM1 magnetic boundary condition is obtained from toroidal ideal MHD plasma response calculations using the GPEC code [31]. GPEC uses a kinetic equilibrium before ELM suppression to calculate the total (plasma + vacuum field) response on the TM1 simulation boundary, resolved into poloidal harmonics $\delta B_{m,n}$ for $m/n$=12/3, 11/3, 10/3 and 9/3. The same kinetic equilibrium profiles taken before ELM suppression, including TRANSP [33] estimates of particle and thermal diffusivities, plasma viscosity, collisionality and resistivity before ELM suppression, are included as initial conditions in the TM1 simulations. TM1 then calculates the resonant field penetration and/or screening for each harmonic and evolves the profiles according to predictions of collisional parallel transport across these islands. For all these ISS plasmas, $D_\perp \approx \chi_\varphi \approx \chi_e \approx 1$ m$^2$/s, with neoclassical resistivity [34] ≈ 1.0×10$^{-7}$ Ωm at the pedestal top from TRANSP analysis.

Fig. 2 shows the TM1 prediction of the $q$=10/3 magnetic island width and resulting electron pressure $P_e$ vs $q_{95}$ for initial profiles obtained from the early ELMing phase of the high-density discharge in Fig. 1a. Fig. 2a shows the evolution of the $q$-profile during the current ramp from $t_1$-$t_5$, while Fig. 2b shows the measured (dotted lines) and TM1 predicted (solid lines) electron pressure profiles for the different $q$-profiles. From Fig. 2b, the height and width of the electron pedestal decrease as $q_{95}$ decreases from $t_1$-$t_4$. The decrease in $q_{95}$ moves the $q$=10/3 resonant surface to out in radius. The island cuts deeper into the pedestal as it moves further out, thus limiting the height and width of the pedestal consistent with experiment. However, at $t_5$ the island moves too far into the pedestal and is screened by the very high E×B and diamagnetic flows in the region of strong pressure gradients. Therefore, at $t_5$ the electron pressure profile jumps back to the initial profile (black dashed curve) in Fig. 2b. The TM1 predicted magnetic island is shown as a function of RMP amplitude and $q_{95}$ in Fig. 2c, and the corresponding pedestal pressure reduction is shown in Fig. 2d. The saturated island width is small ($\Delta\psi_N$≈0.01-0.02) which explains why magnetic islands have not been observed directly using conventional diagnostics in the DIII-D pedestal. The narrow island produces negligible pressure reduction for $q_{95}$≈3.8 (Fig. 2d) because it is too far in from the pedestal top. However, as $q_{95}$ decreases, the island cuts deeper into the pedestal producing a significant reduction in the pedestal pressure $P_{e,ped}$ (Fig. 2d), and a similar reduction in the pedestal width (see profiles in Fig. 2b at $t_3$-$t_4$). Finally, the island is screened (i.e. $t_5$) when the resonant surface moves too far into the pedestal. These simulations are consistent with heuristic arguments originally proposed for how magnetic islands could stabilize ELMs by preventing the expansion of the pedestal to an unstable width [29].

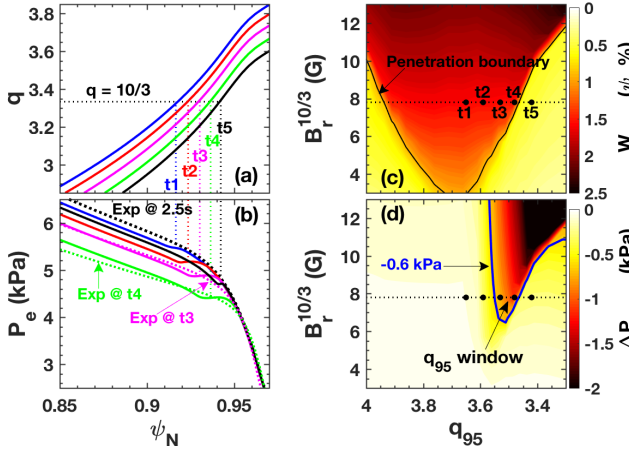

**Fig. 2**. TM1 simulation of the penetration of the $m/n = 10/3$ RMP and its effect on the electron pressure at different $q_{95}$ for #145380: (a) $q$ profiles at different times from Fig. 1d, (b) electron pressure profile from TM1 (solid) vs measurement (dashed) for each $q$-profile from $t_1$-$t_5$, (c) island width $W_{10/3}$ from TM1 and (d) change in pedestal pressure $\Delta P_{e,ped}$ from TM1 versus RMP strength and $q_{95}$, showing the five different times as black dots. The experimental pressure profiles before ELM suppression (black dashed) is shown in Fig. 2b.

Next, the simulations in Fig. 2 are extended over the range of $q_{95}$ in experiment and the evolution of the pedestal pressure is compared with the measurements. Fig. 3a shows the TM1 predicted electron pedestal pressure (red) overlaid with the experimental pressure (blue) vs $q_{95}$ using the GPEC calculated harmonics $m/n$=12/3, 11/3, 10/3 and 9/3. The simulations reproduce the observed reduction in the pedestal pressure during periods of ELM suppression seen in Fig. 1a (and indicated by the yellow bands in Fig. 3a). The weakest pressure reduction with the narrowest $q_{95}$ window is for $m/n$=11/3, corresponding to the sparse ELM window in Fig. 1a. Fig. 3a shows the EPED [29] prediction of $P_{e,ped}$ limited by PBM stability (black line). The EPED model calculates the MHD stability limited height and width of the pedestal at each $q_{95}$ value based on measurements of plasma parameter such as pedestal density, poloidal beta, shape, etc. The TM1 predicted and experimentally measured pedestal pressure are similar and fall significantly below the EPED prediction (≈15%) during the periods of ELM suppression.

At lower pedestal density ($n_{e,ped}$=2.5×10$^{19}$m$^{-3}$ in Fig. 1b), Fig. 3b shows the TM1 prediction of the pedestal pressure (in red), compared to the pedestal pressure (blue) and the EPED prediction (black). The threshold for resonant field penetration drops by ≈25% while the RMP amplitude from GPEC drops by only ≈12% (from 8 to 7 Gauss). Similar to Fig. 3a, TM1 takes the profiles and transport coefficients from the ELMing phase of the discharge ($q_{95}$≈4) to predict the pedestal pressure vs $q_{95}$. An important feature at lower density is that the 11/3 magnetic island produces a stronger pedestal pressure reduction than at higher density in Fig. 3a, leading to robust ELM suppression (yellow band around $q_{95}$≈3.8). From TM1 simulations, the emergence of ELM suppression at $q_{95}$≈11/3 is due to the reduction in the resonant field penetration threshold at lower density. Similarly, the depth and width of the pressure reduction in the 10/3 and 9/3 ELM suppression windows increases, and these begin to merge. To understand why the $q_{95}$ width of ELM suppression increases as the density decreases, we focus our attention on the 10/3 resonance in Fig. 3a,b. We see that the start of the pressure reduction (beginning of the yellow band) is identical for the two discharges, around $q_{95}$=3.6, however the pressure reduction extends to lower $q_{95}$ at lower density. From TM1 the predicted width of the islands are very similar for the two discharges, as the width primarily depends of the RMP amplitude. Thus, the $q_{95}$ where the pressure reduction begins to take place is similar in the two discharges. However, when the threshold for field penetration decreases relative to the applied RMP amplitude, the $m/n$=10/3 island can penetrate deeper into the steep gradient pedestal region (i.e., can exist at lower $q_{95}$) before getting screened. The TM1 prediction in Fig. 3b is consistent with the measured $P_{e,ped}$ (and pedestal width – not shown) and ELMs are suppressed when $P_{e,ped}$ falls significantly (≈15%) below the EPED prediction.

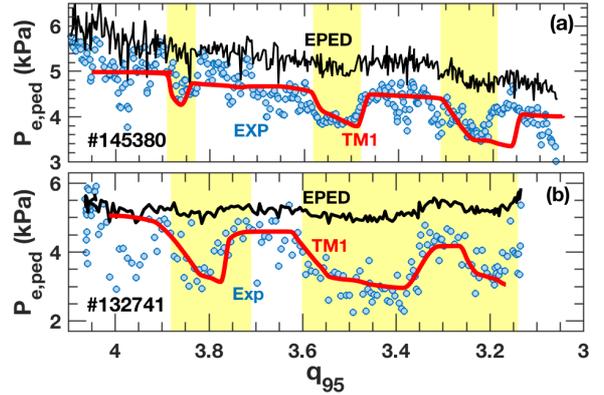

**Fig. 3** Comparison of the pedestal pressure vs $q_{95}$ from experiment (blue), EPED prediction (black) and TM1 simulation (red) at (a) high density (#145380) and (b) intermediate density (#132741).

Finally, at the lowest pedestal density (in Fig. 1c with $n_{e,ped}$≈1.5×10$^{19}$m$^{-3}$) we show that multiple magnetic islands can form on adjacent rational surfaces near the knee of the pedestal, unlike at the higher densities, leading to $q_{95}$ window merger and wide $q_{95}$ ELM suppression windows. Figure 4 shows the TM1 prediction of $P_{e,ped}$ (red), the measured pedestal pressure (blue) and the EPED prediction (black) for the low density discharge in Fig. 1c. The threshold for resonant field penetration drops by ≈50% relative to the highest density discharge in Fig. 1a, based on TM1 simulations, while the RMP amplitude from GPEC drops by only 20% (8 to 6.5 Gauss). Again, TM1 takes the profiles and transport coefficients during the ELMing phase of the discharge and computes the threshold for resonant field penetration and the effect of collisional parallel transport across the islands on the pedestal pressure. From Fig. 4a, TM1 predicts a flat response of the pedestal pressure vs $q_{95}$ consistent with experiment, and a large (≈45%) pedestal pressure reduction relative to EPED. A Poincaré plot of the magnetic field structure vs poloidal magnetic flux in Fig. 4b for $q_{95}$=3.225 shows the presence of two magnetic islands, the smaller 10/3 magnetic island at larger radius and the 9/3 magnetic island near to the top of the pedestal. Overlaid on the Poincare plot is the initial pressure profile (blue) before field penetration and the TM1 predicted pressure profile, which also closely matches experiment.

From TM1 simulations [20], the threshold for field penetration decreases with decreasing density. This allows the 10/3 magnetic island to move deeper into the pedestal before being screening and allows the 9/3 magnetic island to begin cutting into the pedestal before the 10/3 island disappears, leading to continuous ELM suppression.

It is important to highlight that even at the lowest pedestal density, the TM1 simulations show no significant magnetic stochasticity. Furthermore, resonant fields remain strong screening from $\psi_N$=0.96-0.98, consistent with the preservation of the edge transport barrier.

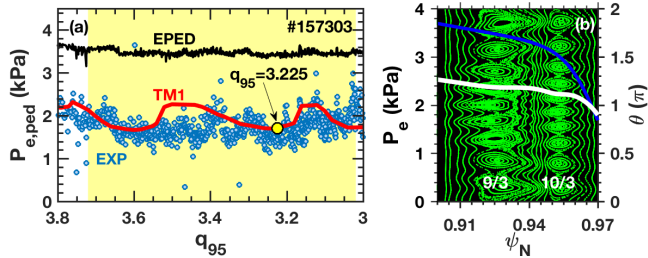

**Fig. 4**. (a) Comparison of the pedestal pressure vs $q_{95}$ from experiment (blue), EPED prediction (black) and TM1 simulation (red) at low plasma density (#157303), and (b) Poincaré plot of the magnetic flux surfaces overlaid with the TM1 predicted pressure profile (white) and original pressure profile (blue) for $q_{95}$=3.225, indicated in Fig. 4a.

It appears from our analysis that wide operational windows of ELM suppression require low pedestal density (and pressure as $T_{e,ped}$≈1keV for all these discharges). However, operating well below the EPED pedestal pressure will not be acceptable in ITER. An important operational question we need to ask is if wide $q_{95}$ windows of ELM suppression can be achieved in DIII-D ISS plasmas at considerably higher pedestal density. We see from Fig. 4 that expanded ELM suppression windows emerge when magnetic islands simultaneously form on adjacent rational surfaces near the top of the pedestal. When the rational surfaces are farther apart then the density (penetration) threshold must be lower for both islands to appear. This is because one of the islands must avoiding getting screened in the steep gradient region of the pedestal. However, if the rational surfaces were to be closer to each other then the density (penetration) threshold can be higher. The distance between the surfaces will decrease as we go to higher toroidal mode number.

We therefore return to the high-density discharge in Fig. 1a and compare the TM1 prediction of the pedestal pressure vs $q_{95}$ for $n$=3 and $n$=4 RMPs. Fig. 5 shows 2D contour plots of the TM1 predicted pedestal pressure reduction vs RMP amplitude and $q_{95}$ for the high-density discharge in Fig. 1a. Fig. 5a is for $n$=3 and Fig. 5b is for $n$=4 RMPs. The $n$=3 harmonics used in TM1 are $m/n$=12/3, 11/3, 10/3, 9/3, while the $n$=4 harmonics are $m/n$=16/4 down to $m/n$=12/4. The solid blue contour in Fig. 5a,b indicates the RMP amplitude required to produce a 15% reduction in $P_{e,ped}$, corresponding to the windows of ELM suppression observed for $n$=3 RMPs in DIII-D (Fig. 3a). It is informative to overlay the 15% pressure reduction contour for the intermediate density case from Fig.1b and 3b (purple dashed line). The $n$=3 RMP amplitude is roughly constant from GPEC for the two discharges in Fig. 5a (horizontal dashed line). The horizontal slice at ≈8 Gauss in Fig. 5a reproduces the observed pedestal pressure reduction shown in Fig. 3a,b. The intermediate density discharge has a lower penetration threshold (bottom of the purple dashed curve vs $q_{95}$) than for the high-density discharge, producing wider ELM suppression windows. In Fig. 5b, the pedestal pressure reduction is calculated for $n$=4 RMPs for the high-density discharge in Fig. 1a and the 15% pressure reduction below the EPED pressure is shown by the blue solid curve.

We see that for an $n$=4 RMP amplitude of ≈ 8 Gauss, similar to the $n$=3 value on the plasma boundary in Fig. 5a, TM1 predicts potentially continuous windows of ELM suppression, but now at high plasma density. These $n$=4 calculations indicate that operationally favorable wide $q_{95}$ ELM suppression windows may be achieved in DIII-D by operating at higher toroidal mode number. The newly planned M-coils [35] for DIII-D can explore this regime, while the planned $n$=3 RMP coils in ITER can actually operate at higher toroidal mode number with the right combination of power supplied and coil connections [36,37].

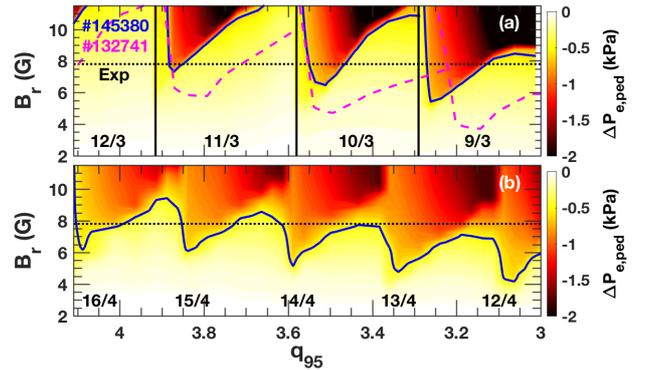

**Fig. 5.** TM1 simulation of the pedestal pressure change $\Delta P_{e,ped}$ versus resonant field strength $B_r$(G) and $q_{95}$ for (a) $n$ = 3 RMP and (b) $n$ = 4 RMP at high plasma density for discharge #145380 (Fig. 1a). The blue solid line corresponds to a 15% reduction in $P_{e,ped}$, and the purple dashed line in Fig. 5a corresponds to a 15% reduction in the pedestal pressure for the intermediate density discharge #132741 (Fig. 1b).

In this Letter we have demonstrated the conditions for accessing wide $q_{95}$ windows of ELM suppression in low-collisionality DIII-D ITER shape plasmas. We have shown using nonlinear two-fluid MHD simulations that the $q_{95}$ range of ELM suppression increases as the threshold for resonant field penetration decreases. For a sufficiently low density, resonant fields can penetrate on more than one rational surface near the top of the pedestal, leading to continuous $q_{95}$ windows of ELM suppression. Our simulations reveal that wide windows of ELM suppression may be accessible at high pedestal pressure in DIII-D using higher toroidal mode number RMPs. The planned M-coil upgrade on DIII-D can explore this regime, with potentially important implications for the operation of the ITER ELM coils. Finally, we cannot rule out that other transport effects may become important at reactor scale. While the two-fluid model is a powerful tool, higher physics fidelity models are also needed to explore the detailed transport physics for extrapolation to reactor scale [38,39].

**Acknowledgements:** Part of the data analysis was performed using the OMFIT integrated modeling framework [40]. This material is based upon work supported by the U.S. Department of Energy, Office of Science, Office of Fusion Energy Sciences, using the DIII-D National Fusion Facility, a DOE Office of Science user facility, under DOE awards DE-AC02-09CH11466 and DOE contract DE-FC02-04ER54698 and DE-FG02-05ER54809.